\documentclass[preprint,twoside,pra,nobibnotes,showpacs,10pt]{revtex4}
\usepackage{amssymb}

\input{tcilatex}

\begin{document}

\title{Nash Equilibrium in the Quantum Battle of Sexes Game}
\author{Jiangfeng Du$^{1,2}$}
\email{djf@ustc.edu.cn}
\author{Xiaodong Xu$^{2}$}
\email{xuxd@mail.ustc.edu.cn}
\author{Hui Li$^{2}$}
\email{lhuy@mail.ustc.edu.cn}
\author{Mingjun Shi$^{1,2}$}
\author{Xianyi Zhou$^{1,2}$}
\author{Rongdian Han$^{1,2}$}
\affiliation{$^{1}$Laboratory of Quantum Communication and Quantum Computation,
University of Science and Technology of China, Hefei, 230026, P.R.China.\\
$^{2}$Department of Modern Physics, University of Science and Technology of
China, Hefei, 230027, P.R.China.}

\begin{abstract}
We investigate Nash Equilibrium in the quantum Battle of Sexes Game. We find
the game has infinite Nash Equilibria and all of them leads to the asymmetry
result. We also show that there is no unique but infinite Nash Equilibrium
in it if we use the quantizing scheme proposed by Eisert et al and the two
players are allowed to adopt any unitary operator as his/her strategies.
\end{abstract}

\maketitle

\section{Introduction}

Game theory is a very useful and important branch of mathematics because it
can solve many problems in economics, social science and biology\cite{1}.
Recently, people are very interested in what would happen when a classical
game is extended into the quantum domain. They did lead to new sight into
the nature of information\cite{2,3,4} and quantum algorithms\cite{5}.Also
some marvelous results are found. Meyer quantized the PQ Game - a coin
tossing game\cite{6} -and found out that one player could increase his
expected payoff by implementing a quantum strategy against his classical
opposite. J.Eisert, M.Wilkens and M.Lewenstein\cite{7} investigated the
Prisoner's Dilemma in the quantum world. They found a unique Nash
Equilibrium, which is different from the classical one, and the dilemma
could be solved if the two players applied quantum strategies and both
players were better satisfied than in the classical world.

Not only 2-player games, but also multi-player games have been shown that
better quantum strategies exist\cite{8}. With all these achievement, it is
natural to think that any classical game has a quantum version with better
solution (payoff) or even think that the quantized game definitely has a
unique Nash Equilibrium, which is better than classical one too.

Recently, Luca Marinatto and Tullio Weber studied the Battle of Sexes Game%
\cite{9}. In that paper they proposed a unique Nash Equilibrium and the
dilemma was moved. However, Benjamin argued that the quantum Battle of the
Sexes game does not in fact have a unique solution\cite{10}. We carefully
study this problem and find out their approach was more like a classical one
with classical possibilities (denoted by ``p'' and ``q''). In this paper, we
first fully quantize this game and find out that the quantum Sexes Game is
much more complicated than the classical one. In this game, there are
virtually infinite Nash Equilibrium rather than unique, and the quantum
payoff is no better than the classical one. Further more, we prove that for
nontrivial 2-player quantum games with the scheme proposed in Ref\cite{7},
if both players can apply any possible pure strategy in the whole SU(2),
there is virtually no Nash Equilibrium.

\section{Quantization of the Battle of Sexes Game}

Our physical model for quantizing the Battle of Sexes Game is presented by
Eisert\cite{7}. Here we depicted it with quantum logic gates showed in
Figure 1.

We send each player a classical 2-state system(a bit) in the zero state. $%
U_{A}$ and $U_{B}$ belong to the strategy space SU(2). We describe it by 
\begin{equation}
U=\left( 
\begin{array}{cc}
e^{i\frac{\varphi +\psi }{2}}\cos \frac{\theta }{2} & ie^{i\frac{\varphi
-\psi }{2}}\sin \frac{\theta }{2} \\ 
ie^{-i\frac{\varphi -\psi }{2}}\sin \frac{\theta }{2} & e^{-i\frac{\varphi
+\psi }{2}}\cos \frac{\theta }{2}%
\end{array}%
\right) ,\ \varphi ,\psi ,\theta \in (-\pi ,\pi )
\end{equation}%
We restrict it with $\psi =\varphi $ , which will be proved reasonable
later, then $U$ should be such as 
\begin{equation}
U=\left( 
\begin{array}{cc}
e^{i\varphi }\cos \frac{\theta }{2} & i\sin \frac{\theta }{2} \\ 
i\sin \frac{\theta }{2} & e^{-i\varphi }\cos \frac{\theta }{2}%
\end{array}%
\right)
\end{equation}%
The payoff matrix for the Battle of sexes game is given as following.

\[
\begin{tabular}{|c|c|c|c|}
\hline
&  & Bob & Bob \\ \hline
&  & O & T \\ \hline
Alice & O & $(\alpha ,\beta )$ & $(\gamma ,\gamma )$ \\ \hline
Alice & T & $(\gamma ,\gamma )$ & $(\beta ,\alpha )$ \\ \hline
\end{tabular}
\]

In the matrix, the first entry in the parenthesis denotes the payoff of
Alice and the second of Bob, and in this game $\alpha >\beta >\gamma $.

We start the game with the initial state $\left| \Psi _{in}\right\rangle
=J\left| OO\right\rangle $, and the outcome state of the process is $\left|
\Psi _{out}\right\rangle =J^{+}U_A\otimes U_BJ\left| OO\right\rangle $.

So the payoff function is

\begin{equation}
\left\{ 
\begin{array}{l}
\$_A=\alpha P_{OO}+\beta P_{TT}+\gamma (P_{OT}+P_{TO}) \\ 
\$_B=\alpha P_{TT}+\beta P_{OO}+\gamma (P_{TO}+P_{OT})%
\end{array}
\right.
\end{equation}
$P_{\sigma \tau }$ is the probability of the different outcome after
measurement. It is described by 
\begin{equation}
P_{\sigma \tau }=\left| \left\langle \sigma \tau \right| J^{+}U_A\otimes
U_BJ\left| OO\right\rangle \right| ^2
\end{equation}
where $\sigma $,$\tau \in \{O,T\}$.

\subsection{Non-Entangled Situation}

The non-entangle condition is $\delta =0$, hence $P_{\sigma \tau }$ can be
expressed by 
\begin{equation}
\left\{ 
\begin{array}{l}
P_{OO}=\cos ^2\frac{\theta _A}2\cos ^2\frac{\theta _B}2 \\ 
P_{TT}=\sin {}^2\frac{\theta _A}2\sin {}^2\frac{\theta _B}2 \\ 
P_{OT}=\cos ^2\frac{\theta _A}2\sin {}^2\frac{\theta _B}2 \\ 
P_{TO}=\sin {}^2\frac{\theta _A}2\cos ^2\frac{\theta _B}2%
\end{array}
\right.
\end{equation}
where A denotes Alice and B denotes Bob.

According to the definition of Nash Equilibrium, we have the following
inequation. 
\begin{equation}
\left\{ 
\begin{array}{l}
\$_A(S_A^{*},S_B^{*})\geqslant \$_A(S_A,S_B^{*}) \\ 
\$_B(S_A^{*},S_B^{*})\geqslant \$_B(S_A^{*},S_B)%
\end{array}
\right. \forall S_A\in SU(2),\forall S_B\in SU(2)
\end{equation}
The * denotes Nash Equilibrium.

We find three profiles of Nash Equilibrium.

1. $\{\theta _A^{*}=\theta _B^{*}=0\}$ . It means that Alice and Bob both
use the identical strategy $\left( 
\begin{array}{cc}
1 & 0 \\ 
0 & 1%
\end{array}
\right) $. The payoff for Alice is $\alpha $ and for Bob is $\beta $ .

2. $\{\theta _A^{*}=\theta _B^{*}=\pi \}$. It means that Alice and Bob both
use the flip operation $i\sigma _x$, which is equivalent to a NOT-Gate. The
payoff for Alice is $\beta $ and for Bob is $\alpha $ .

3. $\{\theta _A^{*}=\arcsin \sqrt{\frac{\beta -\gamma }{\alpha +\beta
-2\gamma }},\theta _B^{*}=\arcsin \sqrt{\frac{\alpha -\gamma }{\alpha +\beta
-2\gamma }}\}$. The payoff is $\$_A=\$_B=\frac{\alpha \beta -\gamma ^2}{%
\alpha +\beta -2\gamma }$.

We can see when $\delta =0$, which is the non-entangle condition, the whole
procedure includes the classical Battle of the Sexes Game and has no novel
characters exceeding it. Also, if $\varphi =0$, but $\delta \neq 0$, the
game appears completely classical and the result is the same as when $\delta
=0$.

\subsection{Max-Entangled Strategy}

If $\delta =\frac \pi 2$, which is the max-entangle condition, and $\varphi
\neq 0$, the situation is quite different from the previous one and the
result is some kind strange.

Now the $P_{\sigma \tau }$'s explicit expression are 
\begin{equation}
\left\{ 
\begin{array}{l}
P_{OO}=\cos ^2\frac{\theta _A}2\cos ^2\frac{\theta _B}2\cos ^2(\varphi
_A+\varphi _B) \\ 
P_{TT}=[\sin \frac{\theta _A}2\sin \frac{\theta _B}2-\cos \frac{\theta _A}%
2\cos \frac{\theta _B}2\sin (\varphi _A+\varphi _B)]^2%
\end{array}
\right.
\end{equation}
because of $P_{TO}+P_{OT}=1-(P_{OO}+P_{TT})$, the payoff is given by 
\begin{equation}
\left\{ 
\begin{array}{l}
\$_A=(\alpha -\gamma )P_{OO}+(\beta -\gamma )P_{TT}+\gamma \\ 
\$_B=(\beta -\gamma )P_{OO}+(\alpha -\gamma )P_{TT}+\gamma%
\end{array}
\right.
\end{equation}

After simply calculation, we find out the strategy profile $\{S_A=S_B=\left( 
\begin{array}{cc}
1 & 0 \\ 
0 & 1%
\end{array}
\right) \}$, which is Nash Equilibrium in the classical game, is no longer
Nash Equilibrium in the quantum field. Since it is easy to see if Bob
unilaterally changes his strategy to $\left( 
\begin{array}{cc}
i & 0 \\ 
0 & -i%
\end{array}
\right) $, because the initial state is $\left| OO\right\rangle $, the
outcome state should be $\left| TT\right\rangle $ . That is to say, Bob can
increase his payoff (form $\beta $ to $\alpha $) by unilaterally changing
his own strategy.

We find infinite Nash Equilibria. Some of them are listed as follows.

\[
\begin{tabular}{|c|c|c|c|c|c|}
\hline
& $\theta _A^{*}$ & $\theta _B^{*}$ & $\varphi _A^{*}+\varphi _B^{*}$ & $%
\$_A $ & $\$_B$ \\ \hline
(1) & $\pi $ & $\pi $ & $\frac \pi 2$ & $\beta $ & $\alpha $ \\ \hline
(2) & $-2\arcsin \sqrt{\frac{\alpha -\beta }{\alpha -\gamma }}$ & $2\arcsin 
\sqrt{\frac{\alpha -\beta }{\alpha -\gamma }}$ & $\frac \pi 2$ & $\beta $ & $%
\alpha $ \\ \hline
(3) & $2\arcsin \sqrt{\frac{\alpha -\beta }{\alpha -\gamma }}$ & $-2\arcsin 
\sqrt{\frac{\alpha -\beta }{\alpha -\gamma }}$ & $\frac \pi 2$ & $\beta $ & $%
\alpha $ \\ \hline
\end{tabular}
\]

The interesting thing is that the payoff for Alice and Bob is always the
same for any different Nash Equilibrium. Actually any Nash Equilibrium leads
to the same asymmetrical result. This is due to the internal asymmetry lying
in the game itself (the Battle of the Sexes Game is an asymmetric game).
Alice always gets $\beta $ and Bob always gets $\alpha $ . From the matrix
we can see that strategy profile (2) and (3) cannot be distinguished.
Further more, they cannot be analogized by the classical counterpart.

\subsection{Reason For Restricted Strategic Space}

Here we prove that in general 2-player quantum games, if the players can
apply any possible pure strategy in the whole SU(2), there is virtually no
Nash Equilibrium. Benjamin\cite{10} pointed out for any $U_A\in SU(2)$,$%
U_B\in SU(2)$, we can always find a matrix $U\in SU(2)$ which satisfies

\begin{equation}
U_A\otimes U_BJ\left| OO\right\rangle =I\otimes (UU_B)J\left| OO\right\rangle
\end{equation}
that is to say, if Alice's strategy is given constant, Bob could make the
outcome state be any eigenstate by unilaterally changing his own strategy.

If the profile $\{S_A^{*},S_B^{*}\}$ is Nash Equilibrium. Because of
equation (8), we can always find an $S_A^{^{\prime }}\in SU(2)$, which
satisfies $\$_A(S_A^{^{\prime }},S_B^{*})=\left( \$_A\right) _{MAX}$, hence
we have $\$_A(S_A^{*},S_B^{*})\geqslant \$_A(S_A^{^{\prime
}},S_B^{*})=\left( \$_A\right) _{MAX}$ (Because of the definition of Nash
Equilibrium).

But in fact, it is always true that $\$_A(S_A^{*},S_B^{*})\leqslant $ $%
\left( \$_A\right) _{MAX}$, hence

\begin{equation}
\$_A(S_A^{*},S_B^{*})=\left( \$_A\right) _{MAX}
\end{equation}

Following the similar process, we can prove

\begin{equation}
\$_B(S_A^{*},S_B^{*})=\left( \$_B\right) _{MAX}
\end{equation}
so

\begin{equation}
\left\{ 
\begin{array}{l}
\left( \$_A\right)
_{MAX}=\$_A(S_A^{*},S_B^{*})=(\$_A)_{00}P_{00}+(\$_A)_{01}P_{01}+(%
\$_A)_{10}P_{10}+(\$_A)_{11}P_{11} \\ 
\left( \$_B\right)
_{MAX}=\$_B(S_A^{*},S_B^{*})=(\$_B)_{00}P_{00}+(\$_B)_{01}P_{01}+(%
\$_B)_{10}P_{10}+(\$_B)_{11}P_{11}%
\end{array}
\right.
\end{equation}
where $P_{\sigma \tau }=\left| \left\langle \sigma \tau \right|
J^{+}U_A\otimes U_BJ\left| OO\right\rangle \right| ^2,$and $\sigma ,\tau $ $%
\in \{0,1\}$.

But for any $\sigma ,\tau $ $\in \{0,1\}$, 
\begin{equation}
(\$_{A})_{\sigma \tau }+(\$_{B})_{\sigma \tau }\leqslant
(\$_{A})_{MAX}+(\$_{B})_{MAX}
\end{equation}%
which is actually four inequations. For a nontrivial game, these inequations
cannot be equations at the same time for any set of $\{\sigma ,\tau \}$, at
least one of them must be ``true'' inequation. Therefore, when we add the
two equations in (12) together with inequation (13), we have

\begin{eqnarray}
(\$_A)_{MAX}+(\$_B)_{MAX}
&<&[(\$_A)_{MAX}+(\$_B)_{MAX}](P_{00}+P_{01}+P_{10}+P_{11})  \nonumber \\
&=&(\$_A)_{MAX}+(\$_B)_{MAX}
\end{eqnarray}
where $P_{00}+P_{01}+P_{10}+P_{11}=1.$ This inequation could never be true.
Hence in general 2-player quantum games, if the players can apply any
possible pure strategy in the whole SU(2), there's virtually no Nash
Equilibrium. This is the reason why we restrict the strategy space to be a
subset of the whole SU(2).

\section{Conclusion}

In conclusion we have fully quantized the Battle of the Sexes Game. The
classical process of the game is included in the non-entanglement situation.
In particular the interesting thing is that, in the entangled situation, we
find infinite Nash equilibrium in the quantum game and the payoff is no
better than it's classical version. We then prove that in most general
2-player static quantum games, there is no Nash equilibrium if both players
can apply any possible pure strategy in the whole SU(2). But this proving is
no longer true while extending into static multi-player quantum games. It
seems that multi-player quantum games may have Nash Equilibrium with the
whole SU(2) as strategy space. So it is interesting to study multi-player
games and the next work is in preparation.

\begin{acknowledgments}
This project was supported by the National Nature Science Foundation of
China and the Science Foundation for Young Scientists of USTC.
\end{acknowledgments}

\section{\textbf{FigureCaption:}}

The setup of 2-player quantum game. Quantum gate $J$ is encoding gate and $%
J^{+}$ is decoding gate, each of them \thinspace is composed of a single
rotation gate and a CNOT gate . Strategy moves of Alice and Bob are
associated with unitary operators $U_{A}$ and $U_{B}$. $J$ is a unitary
operator which is known to both players and symmetric with respect to the
interchange of the two players.

\end{document}